\documentstyle[12pt]{article}
\makeatletter
\def\section{\@startsection {section}{1}{\z@}{3.5ex plus 1ex minus
 .2ex}{2.3ex plus .2ex}{\Large\bf}}
\def\eqnarray{%
 \stepcounter{equation}%
 \let\@currentlabel=\theequation
 \global\@eqnswtrue
 \global\@eqcnt\z@
 \tabskip\@centering
 \let\\=\@eqncr
 $$\halign to \displaywidth\bgroup\@eqnsel\hskip\@centering
 $\displaystyle\tabskip\z@{##}$&\global\@eqcnt\@ne
 \hfil$\displaystyle{{}##{}}$\hfil
 &\global\@eqcnt\tw@$\displaystyle\tabskip\z@{##}$\hfil
 \tabskip\@centering&\llap{##}\tabskip\z@\cr}
\def\maketitle{\par
 \begingroup
 \def\thefootnote{\fnsymbol{footnote}}
 \def\@makefnmark{\hbox
 to 0pt{$^{\@thefnmark}$\hss}}
 \if@twocolumn
 \twocolumn[\@maketitle]
 \else \newpage
 \global\@topnum\z@ \@maketitle \fi\thispagestyle{empty}\@thanks
 \newpage
 \endgroup
 \setcounter{footnote}{0}
 \let\maketitle\relax
 \let\@maketitle\relax
 \gdef\@thanks{}\gdef\@author{}\gdef\@title{}
 \gdef\@prepri{}\gdef\@address{}
 \let\thanks\relax}
\def\@maketitle{\newpage
\begin{flushright}
{\em Ochanomizu University}
\hfill
\@prepri \\
\@date
\end{flushright}
 \vskip 2em \begin{center}
 {\LARGE \@title \par} \vskip 2em
 {\large \lineskip .5em
 \begin{tabular}[t]{c}\@author
 \end{tabular}\par}
 \vskip .5em
 {\em \begin{tabular}[t]{c}\@address
 \end{tabular}\par}
 \end{center}
 \vskip 1.5em
 \vfill
\@ifundefined{@abst}{}{
 \small
 \begin{center}
 {\bf Abstract\vspace{-.5em}\vspace{0pt}}
 \end{center}
 \begin{quotation} \@abst \end{quotation}
 \vfill
 \gdef\@abst{}}}
\long\def\abst#1{\long\gdef\@abst{#1}}
\def\prepri#1{\gdef\@prepri{OCHA-PP-#1}}
\def\address#1{\gdef\@address{#1}}
\makeatother

\begin{document}
\prepri{35}
\date{\today}
\title{Radiative Corrections \\
       in the Presence of \\
       Majorana Neutrinos
       \thanks{This talk was given at the Workshop \lq \lq BDK 93 "
       held at YITP(Kyoto) on June 1993; it is based on the work in
       collaboration with Y. Katsuki, R. Najima, J. Saito and A. Sugamoto.}}
\author{M. Marui}
\address{Department of Physics, Faculty of Science, \\
         Ochanomizu University, \\
         1-1 Otsuka 2,   Bunkyo-ku, Tokyo 112, Japan}
\abst{We have studied the radiative corrections from the fourth generation
      leptons in the context of the see-saw mechanism. We have estimated
      numerically the differential cross section for the process
      (${\rm e}^{+}{\rm e}^{-}\! \rightarrow {\rm W}^{+}{\rm W}^{-}$)
      at one-loop level, and found in the cross section the threshold
      behaviors for Majorana neutrino productions.}
\maketitle
\section{Introduction}
Recent measurements at LEP and SLC do not necessarily rule out the existence
of the fourth generation. What they have established is that the number of
neutrino species whose masses are lighter than $45$ GeV is to be three.
When we consider the fourth generation as a replication of the well-known
generations, the following remarkable feature comes out; the first three
generations have nearly massless neutrinos, while the fourth one has a very
heavy neutrino.

In order to explain the lightness of the observed neutrinos, we usually
introduce the see-saw mechanism \cite{See-Saw}, including right-handed
neutrinos into the first three generations. In this mechanism,
neutrinos acquire a Dirac mass $m_{{\rm D}}$ and a right-handed Majorana
mass $M_{{\rm R}}$, which are reduced to two mass eigenvalues $M_{1}$ and
$M_{2}$ through left-right mixing. We expect that the Dirac masses of
neutrinos generated by the vacuum expectation value of ordinary Higgs bosons
are comparable to masses of their charged partners. On the other hand,
the Majorana mass originates from \lq \lq beyond the standard model".
Therefore this $M_{{\rm R}}$ should be much greater than Dirac masses
$m_{{\rm D}}$, so that we are able to identify eigenvalues $M_{1}$
with observed small masses of neutrinos.

We can also apply this see-saw mechanism to a heavier generation \cite{HP},
which leads to an interesting result as follows:
The LEP experiments suggest that the mass $M_{\nu }$ of the fourth neutrino
is greater than $45$ GeV. This $M_{\nu }$ is identified with the smaller
mass eigenvalue $M_{1}$, where the smaller and larger eigenvalues,
$M_{1}$ and $M_{2}$, are related to the Majorana and Dirac masses through
the relations, $M_{2}-M_{1}=M_{{\rm R}}$ and $M_{1}M_{2}=m_{{\rm D}}^{2}$.
Although $m_{{\rm D}}$ is large, it is bounded above because of the
requirement of triviality; $m_{{\rm D}}\sim m_{{\rm W}}$. Consequently,
it turns out that the Majorana mass $M_{{\rm R}}$ cannot be so large,
and both mass eigenvalues $M_{1}$ and $M_{2}$ are of order $m_{{\rm W}}$,
{\em i.e.}, the see-saw is almost balancing in the heavy neutrino sector.
In this case, the existence of the Majorana neutrinos affects various weak
processes through the large left-right mixing. Thus the contributions from
heavy Majorana neutrinos must be taken into account when analyzing the
next decade experiments whose energy scales are close to $M_{{\rm R}}$.

Here we concentrate on the radiative corrections from these heavy neutrinos
in the process ${\rm e}^{+}{\rm e}^{-}\! \rightarrow {\rm W}^{+}{\rm W}^{-}$
which is the most important process in the LEPII experiments. In the processes
${\rm e}^{+}{\rm e}^{-}\! \rightarrow {\rm f}\overline{{\rm f}}$
observed at the current colliders (LEP, SLC), the effects of heavy neutrinos
appear only in the corrections to the self erergies of gauge bosons
\cite{KL,PT}. Contributions to $S$, $T$ parameters \cite{PT,GR}, which only
self energies are converted into, have already calculated and turned out to
be negative when $M_{1}/m_{{\rm D}}\sim 0.4$ \cite{BS,GT}. We will pay
attention to this Majorana mass region, because precision mesurements favor
negative values for $S$ and $T$ \cite{Neg-ST}. In the W pair production
process, we should take into account radiative corrections not only
on self energies but also trilinear gauge vertices due to the gauge
invariance.

In this talk, we study the one-loop corrections from the fourth generation
leptons in above mentioned LEPII process in the context of see-saw mechanism
with $M_{{\rm R}}$ of order $10^{2}$ GeV. We calculate the differential
cross section numerically and analyze the threshold behavior for the various
types of neutrino production.
\section{Models and amplitudes at one-loop level}
We consider the fourth generation leptons having no mixing with other
generations. Our notation is almost the same as that of ref.\cite{BS,GT}. To
make this talk self-contained, however, we shall exhibit the notational
convention explicitly.

We assume the following mass term for neutrinos
\[
{\cal L}_{{\rm mass}}
  =-\frac{1}{2}
     \overline{(\nu _{{\rm L}}\; \nu _{{\rm R}}^{{\rm c}})}
       \left( \begin{array}{cc}
                0           & m_{{\rm D}} \\
                m_{{\rm D}} & M_{{\rm R}}
              \end{array} \right)
       \left( \begin{array}{c}
                \nu _{{\rm L}}^{{\rm c}} \\ \nu _{{\rm R}}
              \end{array} \right)
     + \mbox{h.c.},
\]
where $\nu ^{{\rm c}}=C\overline{\nu }^{{\rm T}}$ is the charge-conjugated
state of $\nu $. Then we diagonalize the mass matrix and perform the chiral
transformation so as to get positive mass eigenvalues, $M_{1}$ and $M_{2}$.
The result is given by
\[
{\cal L}_{{\rm mass}}
  =-\frac{1}{2}
     \overline{(N_{1}\; N_{2})}
      \left( \begin{array}{cc}
                M_{1} & 0    \\
                0     & M_{2}
             \end{array} \right)
      \left( \begin{array}{c} N_{1} \\ N_{2} \end{array} \right), \\
\]
where
\begin{eqnarray*}
 \left( \begin{array}{c} N_{1} \\ N_{2} \end{array} \right) & = &
  \left( \begin{array}{cc}
            i\gamma _{5}c_{\theta } & -i\gamma _{5}s_{\theta } \\
            s_{\theta }             & c_{\theta }
         \end{array} \right)
  \left( \begin{array}{c}
            \nu _{{\rm L}} + \nu _{{\rm L}}^{{\rm c}} \\
            \nu _{{\rm R}} + \nu _{{\rm R}}^{{\rm c}}
         \end{array} \right), \nonumber \\
 M_{1,2} & = &
  \frac{1}{2}(\sqrt{M_{{\rm R}}^{2}+4m_{{\rm D}}^{2}} \mp M_{{\rm R}}),
\\
 \tan \theta & = & \frac{M_{1}}{m_{{\rm D}}}
               = (\frac{M_{2}}{m_{{\rm D}}})^{-1}.
\end{eqnarray*}
As described in Sec.1, $\tan \theta $ is $\sim O(0.1)$ for the fourth
generation.

Now the gauge interactions of the fourth generation leptons can be
written in terms of mass-eigenstates,
\begin{eqnarray}
 {\cal L}_{{\rm CC}} & = &
   \frac{g}{\sqrt{2}}[W_{+}^{\mu }
    \{ (-ic_{\theta }\overline{N_{1}}+s_{\theta }\overline{N_{2}})
     \gamma _{\mu }\frac{1-\gamma _{5}}{2}E\} \nonumber \\
   & & \qquad \mbox{} + W_{-}^{\mu }
    \{ \overline{E}\gamma _{\mu }\frac{1-\gamma _{5}}{2}
    (ic_{\theta }N_{1}+s_{\theta }N_{2})\} ], \nonumber \\
 {\cal L}_{{\rm NC}} & = &
   -eA^{\mu }\overline{E}\gamma _{\mu }E \nonumber \\
   & &
   \mbox{}+\frac{g}{c_{{\rm W}}}Z^{\mu }
   [\overline{E}\{ (-\frac{1}{4}+s_{W}^{2})\gamma _{\mu }
                    +\frac{1}{4}\gamma _{\mu }\gamma _{5}\} E \nonumber \\
   & & \qquad
    -\frac{1}{4}c_{\theta }^{2}\overline{N_{1}}\gamma _{\mu }\gamma _{5}N_{1}
    -\frac{1}{4}s_{\theta }^{2}\overline{N_{2}}\gamma _{\mu }\gamma _{5}N_{2}
    +\frac{i}{2}s_{\theta }c_{\theta }\overline{N_{2}}\gamma _{\mu }N_{1}],
\end{eqnarray}
where $E$ refers to charged lepton field. We note that the Z coupling to
neutrinos with the same masses is axial vector, while it becomes a vector
coupling when mixing the neutrinos with different masses.

In the ${\rm e}^{+}{\rm e}^{-}\! \rightarrow {\rm W}^{+}{\rm W}^{-}$ reaction,
one-loop corrections carried out using eq.(1) appear both in the self energies
and the trilinear vertices of gauge bosons. The places in which radiative
corrections are necessary are depicted as shaded blobs in fig.1, where the
internal line of neutrino in the t-channel exchange diagram is free from
corrections. We denote the self energies as
$\Pi _{{\rm AA}},\, \Pi _{{\rm AZ}},\, \Pi _{{\rm ZZ}},\, \Pi _{{\rm WW}}$
and the vertex corrections as
$\Gamma _{{\rm AWW}}^{\mu \alpha \beta },\,
\Gamma _{{\rm ZWW}}^{\mu \alpha \beta }$,
\begin{eqnarray*}
\Pi _{{\rm AA}} & = & e_{0}^{2}\Pi _{QQ},
 \qquad \Pi _{{\rm AZ}}
  =\frac{e_{0}^{2}}{s_{0}c_{0}}(\Pi _{3Q}-s_{0}^{2}\Pi _{QQ}),
\\
\Pi _{{\rm ZZ}} & = & (\frac{e_{0}}{s_{0}c_{0}})^{2}
  (\Pi _{33}-2s_{0}^{2}\Pi _{3Q}+s_{0}^{4}\Pi _{QQ}),
 \qquad \Pi _{{\rm WW}}=(\frac{e_{0}^{2}}{s_{0}})^{2}\Pi _{11},
\\
\Gamma _{{\rm AWW}}^{\mu \alpha \beta } & = &
   e_{0}g_{0}^{2}\Sigma _{Q11}^{\mu \alpha \beta },
 \qquad \Gamma _{{\rm ZWW}}^{\mu \alpha \beta }
  =\frac{e_{0}}{s_{0}c_{0}}g_{0}^{2}
   (\Sigma _{311}^{\mu \alpha \beta }
   -s_{0}^{2}\Sigma _{Q11}^{\mu \alpha \beta }),
\end{eqnarray*}
where $s_{0},\, c_{0}$ are defined by $s_{0}=e_{0}/g_{0}$ and
$1\sim 3,\, Q$ refer to vector indices of $SU(2)$ and $U(1)$-charge,
respectively. Possible diagrams corresponding to $\Pi $'s and $\Gamma $'s
are given in fig.2. Generally speaking, radiative corrections are called
\lq \lq oblique" ones \cite{KL,PT} if they appear only on gauge bosons and
not directly on light fermions. The diagrams in fig.2 belong to such oblique
corrections.

A renormalization program for oblique corrections is proposed by Kennedy and
Lynn \cite{KL} on the concept of effective lagrangian. This scheme performed
in the processes
${\rm e}^{+}{\rm e}^{-}\! \rightarrow {\rm f}\overline{{\rm f}}$,
containing only self energies, is familiar to us. In these processes, we can
shuffle all the appearing self energies into six bare parameters, and define
the corresponding running parameters; the coupling constant
($e_{*}$($g_{*}$)), the Weinberg angle ($s_{*}$), the W and Z masses
($M_{{\rm W}*},\, M_{{\rm Z}*}$) and the wave function renormalization
constants of W and Z ($Z_{{\rm W}*},\, Z_{{\rm Z}*}$). Thus the amplitude for
${\rm e}^{+}{\rm e}^{-}\! \rightarrow {\rm f}\overline{{\rm f}}$ at one-loop
level has the same form as the one at the tree level, except that all bare
parameters are replaced by the corresponding \lq \lq starred" parameters.
On the other hand, in the process of
${\rm e}^{+}{\rm e}^{-}\! \rightarrow {\rm W}^{+}{\rm W}^{-}$, which we are
considering here, additional corrections appear in the vertices. The extension
of the formalism of Kennedy and Lynn to this process is given by
Ahn {\em et al.} \cite{APLS}. In this regard, divergences in the vertex
corrections can also be removed without modifying the definitions of the
\lq \lq starred" parameters. The vertex functions at the one-loop level,
however, have additional Lorentz structures not found at tree level.
We can formally write the $s$-channel amplitude for the on-shell W's as
\begin{equation}
{\cal M}=-\frac{ie_{*}^{2}}{P^{2}}Z_{{\rm W}*}\overline{v}\gamma _{\mu }u
         \Gamma ^{\mu \alpha \beta }
         \varepsilon _{\alpha }^{*}(q)\varepsilon _{\beta }^{*}(\overline{q}),
\end{equation}
where $P,\, q,\, \overline{q}$ are momenta as depicted in fig.1, $u$ and $v$
are spinors of electron and positron, and
$\varepsilon _{\alpha }(q),\, \varepsilon _{\beta }(\overline{q})$
refer to polarization-vectors of ${\rm W}^{\mp}$, respectively.
The vertex function $\Gamma ^{\mu \alpha \beta }$ combining A and Z vertices
is expressed in terms of canonical Lorentz structures parametrized by
Hagiwara {\em et al.},
\[
\Gamma ^{\mu \alpha \beta }
 =\sum_{i=1}^{7}(Qf_{i}^{{\rm A}}
  +\frac{I_{3}-s_{*}^{2}Q}{s_{*}^{2}}\frac{P^{2}}{P^{2}-m_{{\rm Z}}^{2}}
  f_{i}^{{\rm Z}})T_{i}^{\mu \alpha \beta },
\]
\begin{eqnarray*}
T_{1}^{\mu \alpha \beta } & = & (q-\overline{q})^{\mu }g^{\alpha \beta },
\qquad \qquad
T_{2}^{\mu \alpha \beta }=\frac{1}{m_{{\rm W}}^{2}}
                          (q-\overline{q})^{\mu }P^{\alpha }P^{\beta },
\\
T_{3}^{\mu \alpha \beta } & = & P^{\alpha }g^{\mu \beta }
                                -P^{\beta }g^{\mu \alpha },
\qquad
T_{4}^{\mu \alpha \beta }=i(P^{\alpha }g^{\mu \beta }
                            +P^{\beta }g^{\mu \alpha }),
\\
T_{5}^{\mu \alpha \beta } & = & \epsilon ^{\mu \alpha \beta \rho }
                                 (q-\overline{q})_{\rho },
\qquad \quad
T_{6}^{\mu \alpha \beta }=i\epsilon ^{\mu \alpha \beta \rho }P_{\rho },
\\
T_{7}^{\mu \alpha \beta } & = &
  \frac{i}{m_{{\rm W}}}(q-\overline{q})^{\mu }
  \epsilon ^{\alpha \beta \rho \sigma }P_{\rho }(q-\overline{q})_{\sigma }.
\end{eqnarray*}
Here, $Q$ and $I_{3}$ are the charge and the isospin of ${\rm e}^{-}$,
respectively. Form factors $f_{i}$ include
$\Delta f_{i(V)}$ and $\Delta f_{i(S)}$ which come from vertex corrections and
self energies,respectively.
\[
f_{i}^{{\rm V=A,Z}}=f_{i(\mbox{tree})}^{{\rm V}}+\Delta f_{i(S)}^{{\rm V}}
                +\Delta f_{i(V)}^{{\rm V}},
\]
\begin{eqnarray}
f_{1(\mbox{tree})}^{{\rm A}} & = & f_{1(\mbox{tree})}^{{\rm Z}}
    =\frac{1}{2}f_{3(\mbox{tree})}^{{\rm A}}
    =\frac{1}{2}f_{3(\mbox{tree})}^{{\rm Z}}=1,
\nonumber \\
f_{i(\mbox{tree})}^{{\rm A}}&= & f_{i(\mbox{tree})}^{{\rm Z}}=0
 \quad (i=\mbox{others}),
\nonumber \\
\Delta f_{1(S)}^{{\rm A}} & = & \frac{1}{2}\Delta f_{3(S)}^{{\rm A}}
    = g_{*}^{2}\frac{\Pi _{3Q}}{P^{2}},
\nonumber \\
\Delta f_{1(S)}^{{\rm Z}} & = & \frac{1}{2}\Delta f_{3(S)}^{{\rm Z}}
    = g_{*}^{2}\frac{\Pi _{3Q}}{P^{2}}
     +\frac{M_{{\rm Z}*}^{2}-m_{{\rm Z}}^{2}}{P^{2}-m_{{\rm z}}^{2}},
\nonumber \\
\Delta f_{i(V)}^{{\rm A}} & = & g_{*}^{2}\Sigma _{Q11}^{i},
\nonumber \\
\Delta f_{i(V)}^{{\rm Z}} & = & \frac{g_{*}^{2}}{c_{*}^{2}}
                                (\Sigma _{311}^{i}-s_{*}^{2}\Sigma _{Q11}^{i}),
\end{eqnarray}
where
\begin{eqnarray*}
M_{{\rm Z}*}^{2}(P^{2})-m_{{\rm Z}}^{2} & = &
 (\frac{e_{*}}{s_{*}c_{*}})^{2}
 [ \{ \Pi _{33}(P^{2})-\Pi _{33}(m_{{\rm Z}})\}
  -\{ \Pi _{3Q}(P^{2})-\Pi _{3Q}(m_{{\rm Z}})\}
\\
& & \mbox{}
  +m_{{\rm Z}}^{2}(c_{*}^{2}-s_{*}^{2})
   \{ \frac{\Pi _{3Q}(P^{2})}{P^{2}}-\frac{\Pi _{3Q}(m_{{\rm Z}})}{P^{2}}\}
\\
& & \mbox{}
  +m_{{\rm Z}}^{2}s_{*}^{4}
   \{ \frac{\Pi _{QQ}(P^{2})}{P^{2}}-\frac{\Pi _{QQ}(m_{{\rm Z}})}{P^{2}}\} ].
\end{eqnarray*}
Above definition of $M_{{\rm Z}*}$ is able to set $Z_{{\rm Z}*}$ to be unity.
Remaining factor $Z_{{\rm W}*}$ in eq.(2) is
\[
Z_{{\rm W}*}(P^{2})=
  1-g_{*}^{2}\{ \frac{\Pi _{3Q}(P^{2})}{P^{2}}-\Pi _{11}'(m_{{\rm W}})\}
\]
This $Z_{{\rm W}*}$ is common to $s$- and $t$-channels.
\section{Results}
Using dimensional regularization, we calculate the one-loop corrections
in eq.(3) from the leptons in the fourth generation. First, we present the
expressions for the self energies, focusing on the threshold behaviors.
\begin{eqnarray}
\lefteqn{16\pi ^{2}\Pi _{33} = }
\hspace*{24pt} \nonumber \\
& &
  \bar{\epsilon }^{-1}
   \{ -\frac{1}{3}P^{2}+\frac{1}{2}M_{{\rm E}}^{2}
      +(c_{\theta }^{4}M_{1}^{2}+s_{\theta }^{4}M_{2}^{2})
      +\frac{1}{2}(M_{1}-M_{2})^{2}\}
\nonumber \\
& & \mbox{}
  -\frac{2}{3}M_{{\rm E}}^{2}
  -\frac{2}{3}(c_{\theta }^{2}M_{1}+s_{\theta }^{2}M_{2})^{2}
  +\frac{s_{\theta }^{2}c_{\theta }^{2}}{3}(M_{1}-M_{2})^{2}
\nonumber \\
& & \mbox{}
  -\frac{1}{2}M_{{\rm E}}^{2}(\ln \frac{M_{{\rm E}}^{2}}{\mu ^{2}}-2)
  -c_{\theta }^{4}M_{1}^{2}(\ln \frac{M_{1}^{2}}{\mu ^{2}}-2)
  -s_{\theta }^{4}M_{2}^{2}(\ln \frac{M_{2}^{2}}{\mu ^{2}}-2)
\nonumber \\
& & \mbox{}
  -\frac{s_{\theta }^{2}c_{\theta }^{2}}{4}(M_{1}^{2}-M_{2}^{2})
   (\ln \frac{M_{1}^{2}}{\mu ^{2}}+\ln \frac{M_{2}^{2}}{\mu ^{2}})
\nonumber \\
& & \mbox{}
  -P^{2}\{ \frac{5}{9}
   -\frac{1}{6}(\ln \frac{M_{{\rm E}}^{2}}{\mu ^{2}}
                +c_{\theta }^{4}\ln \frac{M_{1}^{2}}{\mu ^{2}}
                +s_{\theta }^{4}\ln \frac{M_{2}^{2}}{\mu ^{2}})
   +s_{\theta }^{2}c_{\theta }^{2}
   (\ln \frac{M_{1}^{2}}{\mu ^{2}}+\ln \frac{M_{2}^{2}}{\mu ^{2}})\}
\nonumber \\
& & \mbox{}
  +\frac{s_{\theta }^{2}c_{\theta }^{2}}{6}\frac{1}{P^{2}}(M_{1}^{2}-M_{2}^{2})
   \{ (M_{1}^{2}-M_{2}^{2})+3M_{1}M_{2}\ln \frac{M_{1}^{2}}{M_{2}^{2}}\}
\nonumber \\
& & \mbox{}
  -\frac{s_{\theta }^{2}c_{\theta }^{2}}{12}\frac{1}{(P^{2})^{2}}
   (M_{1}^{2}-M_{2}^{2})^{3}\ln \frac{M_{1}^{2}}{M_{2}^{2}}
\nonumber \\
& & \mbox{}
  -\frac{1}{12}(P^{2}\beta _{{\rm EE}}^{2}+3M_{{\rm E}}^{2})\beta _{{\rm EE}}
      \ln \frac{-P^{2}+2M_{{\rm E}}^{2}+P^{2}\beta _{{\rm EE}}}
               {-P^{2}+2M_{{\rm E}}^{2}-P^{2}\beta _{{\rm EE}}}
\nonumber \\
& & \mbox{}
  -\frac{c^{4}}{12}P^{2}\beta _{{\rm N}_{1}{\rm N}_{1}}^{3}
      \ln \frac{-P^{2}+2M_{1}^{2}+P^{2}\beta _{{\rm N}_{1}{\rm N}_{1}}}
               {-P^{2}+2M_{1}^{2}-P^{2}\beta _{{\rm N}_{1}{\rm N}_{1}}}
\nonumber \\
& & \mbox{}
  -\frac{s^{4}}{12}P^{2}\beta _{{\rm N}_{2}{\rm N}_{2}}^{3}
      \ln \frac{-P^{2}+2M_{2}^{2}+P^{2}\beta _{{\rm N}_{2}{\rm N}_{2}}}
               {-P^{2}+2M_{2}^{2}-P^{2}\beta _{{\rm N}_{2}{\rm N}_{2}}}
\nonumber \\
& & \mbox{}
  -\frac{s^{2}c^{2}}{12}
    \{ 2p^{2}-(M_{1}^{2}+M_{2}^{2}-6M_{1}M_{2})
       -\frac{(M_{1}^{2}-M_{2}^{2})^{2}}{P^{2}}\}
\nonumber \\
& & \qquad \qquad \times
       \beta _{{\rm N}_{1}{\rm N}_{2}}
        \ln \frac{-P^{2}+M_{1}^{2}+M_{2}^{2}
                   +P^{2}\beta _{{\rm N}_{1}{\rm N}_{2}}}
                 {-P^{2}+M_{1}^{2}+M_{2}^{2}
                   -P^{2}\beta _{{\rm N}_{1}{\rm N}_{2}}},
\end{eqnarray}
where $\bar{\epsilon }^{-1}\equiv \epsilon ^{-1}-\gamma + \ln 2\pi $
($\gamma $; Euler's constant), $\mu $ is the reference point, and
$\beta _{ij}$ are defined by,
\[ \beta _{ij}=[1-\frac{(m_{i}+m_{j})^{2}}{P^{2}}]^{1/2}. \]
In eq.(4),
$\beta _{ij} \ln [(-P^{2}+m_{i}^{2}+m_{j}^{2}+P^{2}\beta _{ij})/
                  (-P^{2}+m_{i}^{2}+m_{j}^{2}-P^{2}\beta _{ij})]$
express the singularities at $P^{2}=(m_{i}+m_{j})^{2}$, giving the threshold
behavior for the production of particles with massses $m_{i}$ and $m_{j}$

We note that the threshold behaviors for ${\rm N}_{1}$s' (${\rm N}_{2}$'s)
pair production and ${\rm N}_{1}$-${\rm N}_{2}$ production are different.
The former is as strong as for the Dirac fermion productions ($\sim \beta $).
The latter is as mild as for the scalar productions ($\sim \beta ^{3}$).
The similar expressions for $\Pi _{QQ},\, \Pi _{3Q},\, \Pi _{11}$ are also
obtained. For more details, see ref.\cite{KMNSS}.

Next, we consider the vertex corrections. In our case, there are four
different kinds of Lorentz structures $T_{1},\, T_{2},\, T_{3}$ and $T_{5}$
\cite{GG,HPZH}, because of the $CP$ invariance.
\begin{eqnarray}
\sum _{i=1}^{7}\Delta f_{i(V)}^{{\rm V}}T_{i} & \sim &
 (\overline{\epsilon }^{-1}-1)(T_{1}+2T_{3})+\frac{1}{6}(T_{1}+3T_{3}+T_{5})
\nonumber \\
& & \mbox{}
 +\sum _{i=1,2,3,5}\Delta f_{i(V)}^{{\rm V}\mbox{(finite)}}T_{i},
\end{eqnarray}
where Lorentz indices are omitted. The divergence in the first term in eq.(5)
cancels that existing in the self energy $\Delta f_{1,3(S)}^{{\rm Z}}$
in eq.(3). The Gauge anomalies in the second term should cancel themselves
if we consider the full fourth generation. We perform the estimation of finite
part of the form factors,
$\Delta f_{i(V)}^{{\rm V}\mbox{(finite)}}\, (i=1,\, 2,\, 3,\, 5)$.
For example, $\Delta f_{1(V)}^{{\rm Z}\mbox{(finite)}}$ in eq.(5) is obtained
explicitly
as follows,
\begin{eqnarray}
\lefteqn{(4\pi )^{2}\frac{c_{*}^{2}}{g_{*}^{2}}
\Delta f_{1}^{{\rm Z}\mbox{(finite)}} = }
\hspace*{24pt} \nonumber \\
& &
\frac{35}{36}c_{*}^{2}
  -\frac{1}{3}(\frac{1}{2}-s_{*}^{2})\ln \frac{M_{{\rm E}}^{2}}{\mu ^{2}}
  -\frac{1}{6}c_{\theta }^{2}\ln \frac{M_{1}^{2}}{\mu ^{2}}
  -\frac{1}{6}s_{\theta }^{2}\ln \frac{M_{2}^{2}}{\mu ^{2}}
\nonumber \\
& & \mbox{}
  -\frac{1}{P^{2}}
     \{ \frac{2}{3}(\frac{1}{2}-s_{*}^{2})M_{{\rm E}}^{2}
       +\frac{1}{3}(c_{\theta }^{2}M_{1}^{2}+s_{\theta }^{2}M_{2}^{2})
       +\frac{1}{4}s_{\theta }^{2}c_{\theta }^{2}(M_{2}^{2}-M_{1}^{2})
           \ln \frac{M_{2}^{2}}{M_{1}^{2}}\}
\nonumber \\
& & \mbox{}
  +\frac{s_{\theta }^{2}c_{\theta }^{2}}{(P^{2})^{2}}
     \{ \frac{1}{3}(M_{2}^{2}-M_{1}^{2})^{2}
       +\frac{1}{4}(M_{2}^{2}-M_{1}^{2})(M_{2}^{2}+M_{1}^{2})
           \ln \frac{M_{2}^{2}}{M_{1}^{2}}\}
\nonumber \\
& & \mbox{}
  -\frac{s_{\theta }^{2}c_{\theta }^{2}}{6}\frac{1}{(P^{2})^{3}}
    (M_{2}^{2}-M_{1}^{2})^{3} \ln \frac{M_{2}^{2}}{M_{1}^{2}}
\nonumber \\
& & \mbox{}
  -\frac{1-2c_{*}^{2}}{12}\frac{P^{2}-M_{{\rm E}}^{2}}{P^{2}}
    \beta _{{\rm EE}}
    \ln \frac{-P^{2}+2M_{{\rm E}}^{2}+P^{2}\beta _{{\rm EE}}}
             {-P^{2}+2M_{{\rm E}}^{2}-P^{2}\beta _{{\rm EE}}}
\nonumber \\
& & \mbox{}
  +\frac{c_{\theta }^{4}}{12}\frac{P^{2}-M_{1}^{2}}{P^{2}}
    \beta _{{\rm N}_{1}{\rm N}_{1}}
    \ln \frac{-P^{2}+2M_{1}^{2}+P^{2}\beta _{{\rm N}_{1}{\rm N}_{1}}}
             {-P^{2}+2M_{1}^{2}-P^{2}\beta _{{\rm N}_{1}{\rm N}_{1}}}
\nonumber \\
& & \mbox{}
  +\frac{s_{\theta }^{4}}{12}\frac{P^{2}-M_{2}^{2}}{P^{2}}
    \beta _{{\rm N}_{2}{\rm N}_{2}}
    \ln \frac{-P^{2}+2M_{2}^{2}+P^{2}\beta _{{\rm N}_{2}{\rm N}_{2}}}
             {-P^{2}+2M_{2}^{2}-P^{2}\beta _{{\rm N}_{2}{\rm N}_{2}}}
\nonumber \\
& & \mbox{}
  +\frac{s_{\theta }^{2}c_{\theta }^{2}}{12}
    \frac{2(P^{2})^{2}-P^{2}(M_{1}^{2}+M_{2}^{2})+2(M_{1}^{2}-M_{2}^{2})^{2}}
         {p^{2}}
\nonumber \\
& & \qquad \qquad \times
       \beta _{{\rm N}_{1}{\rm N}_{2}}
        \ln \frac{-P^{2}+M_{1}^{2}+M_{2}^{2}
                   +P^{2}\beta _{{\rm N}_{1}{\rm N}_{2}}}
                 {-P^{2}+M_{1}^{2}+M_{2}^{2}
                   -P^{2}\beta _{{\rm N}_{1}{\rm N}_{2}}},
\nonumber \\
& & \mbox{}
  +c_{\theta }^{2}\int \! dxdy\,
   \frac{(\frac{1}{2}-s_{*}^{2})N(x,y;M_{{\rm E}},M_{{\rm E}},M_{1})
         -\frac{1}{2}s_{*}^{2}M_{{\rm E}}^{2}(x+y)}
        {D(x,y;M_{{\rm E}},M_{{\rm E}},M_{1})}
\nonumber \\
& & \mbox{}
  +s_{\theta }^{2}\int \! dxdy\,
   \frac{(\frac{1}{2}-s_{*}^{2})N(x,y;M_{{\rm E}},M_{{\rm E}},M_{2})
         -\frac{1}{2}s_{*}^{2}M_{{\rm E}}^{2}(x+y)}
        {D(x,y;M_{{\rm E}},M_{{\rm E}},M_{2})}
\nonumber \\
& & \mbox{}
  +\frac{1}{2}c_{\theta }^{4}\int \! dxdy\,
   \frac{N(x,y;M_{1},M_{1},M_{{\rm E}})-\frac{1}{2}M_{1}^{2}(x+y)}
        {D(x,y;M_{1},M_{1},M_{{\rm E}})}
\nonumber \\
& & \mbox{}
  +\frac{1}{2}s_{\theta }^{4}\int \! dxdy\,
   \frac{N(x,y;M_{2},M_{2},M_{{\rm E}})-\frac{1}{2}M_{2}^{2}(x+y)}
        {D(x,y;M_{2},M_{2},M_{{\rm E}})}
\nonumber \\
& & \mbox{}
  +\frac{1}{2}s_{\theta }^{2}c_{\theta }^{2}\int \! dxdy\,
   \frac{N(x,y;M_{1},M_{2},M_{{\rm E}})+\frac{1}{2}M_{1}M_{2}(x+y)}
        {D(x,y;M_{1},M_{2},M_{{\rm E}})},
\end{eqnarray}
where
\begin{eqnarray*}
N(x,y;m_{a},m_{b},m_{c}) & = &
    -\frac{1}{2}m_{{\rm W}}^{2}(x^{3}-x^{2}y-xy^{2}+y^{3}-x^{2}-y^{2}+x+y) \\
& & \mbox{}
    +\frac{1}{2}P^{2}xy
    +\frac{1}{2}(-m_{a}^{2}-m_{b}^{2}+2m_{c}^{2})xy-\frac{1}{2}m_{c}^{2}, \\
D(x,y;m_{a},m_{b},m_{c}) & = &
    -m_{{\rm W}}^{2}(x+y)(1-x-y)-P^{2}xy \\
& & \mbox{}
    +m_{a}^{2}x+m_{b}^{2}y+m_{c}^{2}(1-x-y).
\end{eqnarray*}

The difference of threshold behaviors between ${\rm N}_{1}$-${\rm N}_{1}$
(${\rm N}_{2}$-${\rm N}_{2}$)
and ${\rm N}_{1}$-${\rm N}_{2}$ productions are contained in the $5\sim 14$
lines of eq.(6). To clarify this difference and analyze observability of
the neutrinos' radiative corrections in our model, we need the numerical
calculations. We get the expressions of the remaining form factors, similarly.
The form factors
$\Delta f_{1(V)}^{{\rm A}},\, \Delta f_{3(V)}^{{\rm A}},\,
\Delta f_{3(V)}^{{\rm Z}}$
have also divergences which cancel in the similar way as in the case of
$\Delta f_{1(V)}^{{\rm Z}}$,
while $\Delta f_{2(V)}^{{\rm A,Z}},\, \Delta f_{5(V)}^{{\rm A,Z}}$ are finite.
Detailed expressions for $\Delta f_{i}$ will be given in ref.\cite{KMNSS}.

Now, we present the numerical results of the differential cross section for
${\rm e}^{+}{\rm e}^{-}\! \rightarrow {\rm W}^{+}{\rm W}^{-}$. We use
numerical method by Fujimoto {\em et al.} \cite{FSKO} to perform double
Feynman parameter integrals in $\Delta f_{i(V)}$. In this analysis, we take
the following values at $m_{{\rm Z}}$ for fixing the parameters,
\begin{eqnarray*}
\frac{4\pi }{e_{*}^{2}} & = & 128.0, \qquad s_{*}^{2} = 0.223, \\
m_{{\rm Z}} & = & 93.00\; \mbox{GeV}, \qquad m_{{\rm W}} = 81.97\; \mbox{GeV},
\end{eqnarray*}
where we neglect the influence of the running of $e_{*},\, s_{*}$
\cite{KL,PT}. We also neglect gauge anomalies. The differential cross section
$d\sigma /d\cos \theta $ for
${\rm e}^{+}{\rm e}^{-}\! \rightarrow
 {\rm W}^{+}_{{\rm L}}{\rm W}_{{\rm L}}^{-}$
at scattering angle $\theta = \frac{\pi }{2}$ are shown in fig.3.
In the process for W's polarized longitudinally, we expect that the radiative
corrections from heavy particles will be enhanced \cite{APLS}.
The dashed curve shows $d\sigma /d\cos \theta $ at tree level, and the solid
curve at one-loop level assuming $M_{{\rm D}}=500\, {\rm GeV}$ and
$\tan \theta =0.4$; S and T parameters are negative for this region.
In comparison, the dotted curve is shown, assuming vanishing Majorana mass,
{\em i.e.}, this is the Dirac neutrino limit of $M_{{\rm D}}=200$ GeV.
Here, these cross sections are expressed in units of the point cross section
$1R=4\pi \alpha ^{2}/3s$ where $s=P^{2}$.

In Dirac neutrino case, we can see peaks at the pair production thresholds of
charged leptons and neutrinos ($\sqrt{s}=400,\, 1000$ GeV), in the similar
manner as in the heavy quark case \cite{APLS}. Heavy leptons, however, bring
small effects because of having no color factor $3$. Especially, the effect
of Dirac neutrinos is very small, since they have no couplings with photons.
For Majorana neutrinos, there is a quite small peak only at
${\rm N}_{1}$-${\rm N}_{2}$ threshold ($\sqrt{s}=M_{1}+M_{2}=1450$ GeV)
because neutrino effects are suppressed due to the small couplings including
mixing angle in eq.(1). We can see no other peaks except for the charged
leptons' pair production threshold. As mentioned above, there exists the
difference of threshold behaviors in self energies, depending on the neutrino
types. Numerical calculations including vertex corrections also suggest that
threshold behaviors are scalar-like for ${\rm N}_{1}$'s (${\rm N}_{2}$'s)
pair production, but Dirac fermion-like for ${\rm N}_{1}$-${\rm N}_{2}$
production. This is caused by the difference in the coupling types between
$Z\overline{N_{1}}N_{1}$ ($Z\overline{N_{2}}N_{2}$) and
$Z\overline{N_{1}}N_{2}$, which can be seen in eq.(1).
\section{Conclusion}
In this talk, we have investigated the one-loop corrections from the heavy
Majorana neutrinos in the differential cross section for the process
${\rm e}^{+}{\rm e}^{-}\! \rightarrow{\rm W}^{+}{\rm W}^{-}$, which will be
seen in the near future at LEPII. We have adopted the formalism of Kennedy
and Lynn as a renormalization procedure, using \lq \lq starred" parameters.
The differential cross section has been estimated numerically, and we have
found that there is no visible peak at the threshold of producing a pair of
light neutrinos (${\rm N}_{1}$-${\rm N}_{1}$) or a pair of heavy neutrinos
(${\rm N}_{2}$-${\rm N}_{2}$). We have recognized that there exists a small
peak at the threshold of producing a pair of light and heavy neutrinos
(${\rm N}_{1}$-${\rm N}_{2}$). Then the effect from the Majorana neutrinos
on the differential cross section at the scattering angle
$\theta = \frac{\pi }{2}$ is quite small at the LEP region. Therefore in order
to investigate further the role of Majorana neutrinos in the weak processes,
we need to study the angular distribution of the differential cross section,
electric (magnetic) moment of W \cite{HPZH,W-Mom}, or some other quantities
which are sensitive to the new physics.
\section*{Acknowledgments}
I am deeply thankful to K. Kato who offered the computer program performing
Feynman Parameter integrals. I would like to thank M. Morikawa and M. Suzuki
for their advice about using the computers. I am also grateful to the
organizers of the workshop for their warm hospitality.

\end{document}